\documentclass[reprint,twocolumn]{revtex4}
\usepackage{amsfonts}
\usepackage{amsmath}
\usepackage{amssymb}
\usepackage{subfigure}
\usepackage{charter}
\usepackage{graphicx}

\begin{document}

\title{Operator transpose within normal ordering and its applications for
quantifying entanglement}
\author{Liyun Hu$^{\dag 1}$, Luping Zhang$^{1}$, Xiaoting Chen$^{1},$ Wei Ye$%
^{1}$, Qin Guo$^{1}$, and Hongyi Fan$^{2}$}
\affiliation{$^{1}${\small Center for Quantum Science and Technology, Jiangxi Normal
University, Nanchang 330022, China}\\
$^{2}${\small Department of Material Science and Engineering, University of}%
\\
{\small Science and Technology of China, Hefei 230026, China}\\
$^{\dag }${\small Corresponding author: hlyun2008@126.com}}

\begin{abstract}
Partial transpose is an important operation for quantifying the
entanglement, here we study the (partial) transpose of any single (two-mode)
operators. Using the Fock-basis expansion, it is found that the transposed
operator of an arbitrary operator can be obtained by replacement of $a^{\dag
}(a)$ by $a(a^{\dag })$ instead of c-number within normal ordering form. The
transpose of displacement operator and Wigner operator are studied, from
which the relation of Wigner function, characteristics function and average
values such as covariance matrix are constructed between density operator
and transposed density operator. These observations can be further extended
to multi-mode cases. As applications, the partial transpose of two-mode
squeezed operator and the entanglement of two-mode squeezed vacuum through a
laser channel are considered.

\end{abstract}

\maketitle

\section{Introduction}

Quantum entanglement is an important resources for achieving different
quantum tasks, including quantum teleportation, quantum key distribution,
quantum dense code, and quantum metrology \cite{1}. Especially for quantum
systems with continuous-variable, how to justify the existence of quantum
entanglement has been a challenging problem. In order to solve this issue,
many entanglement criteria and measure methods are proposed \cite%
{2,3,4,5,6,7,8,9,10,11,12,13}, such as Peres-Horodecki criterion, Simon
criterion, Shchukin and Vogel criterion, Hillery and Zubairy criteria,
Agarwal and Biswas criteria as well as Nha and Kim criteria, and entropy of
entanglement, entanglement of formation, as well as logarithmic negativity.
For instance, the Simon criterion is turned out to be a necessary and
sufficient condition for all bipartite Gaussian states \cite{2}, but it
cannot be used to measure the entanglement of non-Gaussian states such as
entangled coherent states \cite{3}. Among these above methods, the
logarithmic negativity is considered as a computable measure of entanglement
for arbitrary bipartite system including both Gaussian and non-Gaussian
states \cite{11}.

In fact, in these criteria or measures, the key starting point is that the
partial transpose on the density operator of two-mode entangled states,
where the resulting operator may not be qualified quantum states whose
eigenvalues are non-negative real numbers. The logarithmic negativity is
proposed based on the negativity of partial transposition, i.e., the
negative eigenvalues. In order to deal with the partial transpose or
calculate the logarithmic negativity, on one hand, it is common method to
appeal to special basis (say Fock basis) or corresponding matrix in selected
basis. On the other hand, for quantum systems with continuous variable, the
method in phase space is also used to discuss the entanglement through the
partial transposition. For instance, Simon extended the Peres-Horodecki
criterion to bipartite continuous variable states from the view point of
phase space \cite{2}. It is shown that the partial transpose actually
corresponds to a mirror reflection, which can be clearly seen by using the
definition of Wigner function in coordinate representation. These above two
methods can be actually considered as indirect ones for discussions about
entanglement or negativity.

It is interesting to notice that, to treat the issue of quantum decoherence
in open system which is described by the master equation, two different ways
have been developed and used widely. One is to convert the master equation
to the equation of distribution function in phase space, such as P-function,
Wigner function, then derive their input-output relations \cite{14,15,16}.
The other is to directly solve the master equation, i.e., to derive the sum
representation of Kraus operator from the viewpoint of operator \cite{17,18}%
. Recently, the Kraus representation has been used to deal with the
precision of quantum metrology in presence of noise \cite{19,20,21}. It will
be beneficial for further revealing the nature and eigen spectrum of the
operator. Motivated by these above ideas, an interesting question arises
naturally: how to directly achieve the transpose or partial transpose of any
operator including density operator from the viewpoint of operator?

In this paper, starting from the definition of operator transpose in Fock
space, we will examine the transpose or partial transpose of any operator
within normal ordering form. It is shown that the (partially) transposed
operator can be obtained by replacing creation and annihilation operators
with annihilation and creation ones within normal ordering form,
respectively, while classical numbers are kept unchanged. Here we should
emphasize that the replacement must be made within normal ordering form
especially for partial transpose case. Then based on this point, we further
examine the relations between operators, distribution functions in phase
space, and average values of operators, before and after the transpose
operation. These discussions can be directly extended to multi-mode cases.
In particular, as an effective and direct application, we analytically
calculate the entanglement degree of the two-mode squeezed vacuum (TMSV)
when passing through the laser channel, by using the logarithmic negativity.

This paper is arranged as follows. In Sec. 2, we give a definition of
operator transpose in Fock space, and then derive a lemma of operator
transpose for normally ordering operator. The transpose of some operators
are discussed, including displacement operator, Wigner operator, and
single-mode squeezing operator. In addition, the relations about
distribution functions and average values are derived before and after the
transpose. This case is extended to two-mode case by taking partial
transpose in Sec. 3. Sec. 4 is devoting to deriving the entanglement degree
of the TMSV through the laser channel. Our conclusions are drawn in Sec. 5.

\section{Transpose of operator}

In this section, we introduce the definition of operator transpose. First,
we consider a simple case, i.e., single-mode operator transpose, including
displacement operator, Wigner operator, and density operator. For any
single-mode operator $\hat{O}$, it can be expanded in Fock space as%
\begin{equation}
\hat{O}=\sum_{j,k}O_{jk}\left\vert j\right\rangle \left\langle k\right\vert
,O_{jk}=\left\langle j\right\vert \hat{O}\left\vert k\right\rangle ,
\label{t1}
\end{equation}%
its transpose, denoted as $\hat{O}^{T}$, can be defined as%
\begin{equation}
\hat{O}^{T}=\sum_{j,k}O_{jk}\left( \left\vert j\right\rangle \left\langle
k\right\vert \right) ^{T}=\sum_{j,k}O_{jk}\left\vert k\right\rangle
\left\langle j\right\vert ,  \label{t2}
\end{equation}%
or%
\begin{equation}
\hat{O}^{T}=\sum_{j,k}O_{kj}\left\vert j\right\rangle \left\langle
k\right\vert .  \label{t3}
\end{equation}%
Here we should notice that the defined transpose for operator is basis
dependent. For different basis such as coherent state basis, the
corresponding results will be different from each other.

\subsection{Transpose of any normally ordering operator}

Here we consider a lemma: for any normally ordering operator, its transpose
can be obtained by replacing $a(a^{\dag})$ with $a^{\dag}(a)$ within its
normal ordering form, i.e,
\begin{equation}
\left[ \colon F\left( a,a^{\dag}\right) \colon \right] ^{T}=\colon F\left(
a^{\dag},a\right) \colon,  \label{t3a}
\end{equation}
where the symbol $\colon \colon$ denotes the normal ordering. The form of
function within normal ordering is kept unchanged.

Proof of Eq. (\ref{t3a}): Expanding the operator $\colon F\left( a,a^{\dag
}\right) \colon$ in Fock space, we have
\begin{equation}
\colon F\left( a,a^{\dag}\right) \colon=\sum_{j,k}\left \vert j\right
\rangle \left \langle j\right \vert \colon F\left( a,a^{\dag}\right) \colon
\left \vert k\right \rangle \left \langle k\right \vert .  \label{t3b}
\end{equation}
Using the coherent state representation of Fock state, i.e., $\left \vert
k\right \rangle =\left. \frac{1}{\sqrt{k!}}\frac{\partial^{k}}{\partial
\tau^{k}}\left \vert \tau \right \rangle \right \vert _{\tau=0}$ in which $%
\left \vert \tau \right \rangle =e^{\tau a^{\dag}}\left \vert
0\right
\rangle $ is un-normalized coherent state \cite{22}, one can see
that
\begin{align}
& \left \langle j\right \vert \colon F\left( a,a^{\dag}\right) \colon \left
\vert k\right \rangle  \notag \\
& =\left. \frac{1}{\sqrt{j!k!}}\frac{\partial^{j+k}}{\partial t^{j}\partial
\tau^{k}}\left \langle t\right \vert \colon F\left( a,a^{\dag}\right) \colon
\left \vert \tau \right \rangle \right \vert _{t=\tau=0}  \notag \\
& =\left. \frac{1}{\sqrt{j!k!}}\frac{\partial^{j+k}}{\partial t^{j}\partial
\tau^{k}}\left \langle \tau \right \vert \colon F\left( a^{\dag },a\right)
\colon \left \vert t\right \rangle \right \vert _{t=\tau=0}  \notag \\
& =\left \langle k\right \vert \colon F\left( a,a^{\dag}\right) \colon \left
\vert j\right \rangle ,  \label{t3c}
\end{align}
thus substituting Eq. (\ref{t3c}) into Eq. (\ref{t3b}) and using the
definition in Eq. (\ref{t2}), it is ready to have
\begin{align}
& \left[ \colon F\left( a,a^{\dag}\right) \colon \right] ^{T}  \notag \\
& =\sum_{j,k}\left \vert k\right \rangle \left \langle k\right \vert \colon
F\left( a,a^{\dag}\right) \colon \left \vert j\right \rangle \left \langle
j\right \vert  \notag \\
& =\colon F\left( a^{\dag},a\right) \colon,  \label{t28b}
\end{align}
which indicates that under the normal ordering form, the transposed operator
can be obtained by making a replacement with $a\leftrightarrow a^{\dag}$
instead of c-number. For a simple instance, for $\left( a^{\dag2}a\right)
^{T}=\left( \colon a^{\dag2}a\colon \right) ^{T}=\colon
a^{2}a^{\dag}\colon=a^{\dag}a^{2}.$

Using this property in Eq. (\ref{t3a}), it is ready to get the transpose of
displacement operator
\begin{equation}
\left[ D\left( \alpha \right) \right] ^{T}=e^{\alpha a-\alpha^{\ast
}a^{\dagger}}=D\left( -\alpha^{\ast}\right) ,  \label{t29}
\end{equation}
and the transpose of Wigner operator \cite{23,24}%
\begin{align}
\left[ \Delta \left( \alpha,\alpha^{\ast}\right) \right] ^{T} & =\frac {1}{%
\pi}\colon \exp \left \{ -2(a-\alpha^{\ast})(a^{\dagger}-\alpha)\right \}
\colon  \label{t29a} \\
& =\Delta \left( \alpha^{\ast},\alpha \right) .  \label{t30}
\end{align}
Eq. (\ref{t29}) indicates that the transpose of displacement operator can be
obtained by replacing $\alpha \ $with$-\alpha^{\ast},$ or
\begin{equation}
D\left( q,p\right) \overset{q\leftrightarrow-q}{\longrightarrow}D\left(
-q,p\right) =\left[ D\left( q,p\right) \right] ^{T},  \label{t10}
\end{equation}
where $\alpha=\left( q+ip\right) /\sqrt{2}$ and $D\left( q,p\right) =D\left(
\alpha \right) =\exp \left \{ i\left( pQ-qP\right) \right \} $ with $%
Q=\left( a+a^{\dagger}\right) /\sqrt{2}$ and $P=(a-a^{\dagger})/\sqrt{2}i.$
While the transpose of Wigner operator can be gotten by replacing $\alpha
\leftrightarrow \alpha^{\ast}$ or $p\leftrightarrow-p$ which is different
from the case of displacement. In addition, the transposed Wigner operator $%
\left[ \Delta \left( \alpha,\alpha^{\ast}\right) \right] ^{T}$ is still an
Hermite operator due to
\begin{equation}
\left[ \Delta \left( \alpha^{\ast},\alpha \right) \right] ^{\dagger}=\Delta
\left( \alpha^{\ast},\alpha \right) ,  \label{t29b}
\end{equation}
which\ can be clearly seen from Eq. (\ref{t29a}). Actually, Eq. (\ref{t30})
can also be derived from Eqs. (\ref{t29a}) and (\ref{t2}). It is interesting
to notice that single-mode squeezing operator $S_{1}(\xi)=\exp[\frac{1}{2}%
\left( \xi a^{\dagger2}-\xi^{\ast}a^{2}\right) ]$ with $\xi=re^{i\theta}$,
whose normally ordering form is given by \cite{25}%
\begin{align}
S_{1}(\xi) & =\text{sech}^{1/2}r\colon \exp \left \{ \frac{1}{2}\left(
e^{i\theta}a^{\dagger2}-e^{-i\theta}a^{2}\right) \tanh r\right \}  \notag \\
& \times \exp \left \{ \left( \text{sech}r-1\right) a^{\dag}a\right \}
\colon,  \label{t29d}
\end{align}
thus according to Eq. (\ref{t3a}), the transpose of $S_{1}(\xi)$ is%
\begin{equation}
\left[ S_{1}(\xi)\right] ^{T}=\exp \left \{ \frac{1}{2}\left( \xi
a^{2}-\xi^{\ast}a^{\dagger2}\right) \right \} =S_{1}(-\xi^{\ast}).
\label{t29e}
\end{equation}
It is easy to see that for complex parameter $\xi$, $\left[ S_{1}(\xi)\right]
^{T}=S_{1}(-\xi^{\ast})\neq S_{1}(-\xi)=[S_{1}(\xi)]^{\dagger}.$ In
particular, for real parameter $\xi$, $\left[ S_{1}(\xi)\right]
^{T}=[S_{1}(\xi)]^{\dagger}$. This case is not true for two-mode correlated
operation, see Eq. (\ref{t47}) below.

\subsection{Characteristics function and Wigner function of transposed
density operator}

\subsubsection{Characteristics function of transposed density operator}

For any single-mode density operator $\rho$, which can be expanded according
to the displacement operator, i.e., the Weyl representation of density
operator \cite{26}%
\begin{equation}
\rho=\int_{-\infty}^{\infty}\frac{d^{2}\alpha}{\pi}\chi \left( \alpha
,\alpha^{\ast}\right) D\left( -\alpha \right) ,  \label{t11}
\end{equation}
or%
\begin{equation}
\rho=\int_{-\infty}^{\infty}\frac{dqdp}{2\pi}\chi \left( q,p\right) D\left(
-q,-p\right) ,  \label{t12}
\end{equation}
thus using Eq. (\ref{t29}), the transpose of density operator $\rho$ is
given by%
\begin{align}
\rho^{T} & =\int_{-\infty}^{\infty}\frac{d^{2}\alpha}{\pi}\chi \left(
\alpha,\alpha^{\ast}\right) D\left( \alpha^{\ast}\right)  \notag \\
& =\int_{-\infty}^{\infty}\frac{d^{2}\alpha}{\pi}\chi \left( -\alpha^{\ast
},-\alpha \right) D\left( -\alpha \right) .  \label{t13}
\end{align}
Eq. (\ref{t13}) indicates the characteristic function of transposed density
operator is given by replacing $\alpha \ $with $-\alpha^{\ast}$, which is
similar to the case of displacement operator, i.e., after the transpose for
density operator, we have
\begin{align}
\chi \left( \alpha,\alpha^{\ast}\right) & \rightarrow \chi \left(
-\alpha^{\ast},-\alpha \right) ,  \label{t14} \\
\chi \left( q,p\right) & \rightarrow \chi \left( -q,p\right) .  \label{t15}
\end{align}

\subsubsection{\textit{Wigner function of transposed density operator}}

Next, we examine the corresponding Wigner function after the transpose of
density operator. For this purpose, we can expand single-mode density
operator $\rho$ by using Wigner operator $\Delta \left( q,p\right) $ \cite%
{23,24}, i.e.,
\begin{equation}
\rho=2\pi \int_{-\infty}^{\infty}W\left( q,p\right) \Delta \left( q,p\right)
dqdp,  \label{t16}
\end{equation}
or%
\begin{equation}
\rho=4\pi \int_{-\infty}^{\infty}W\left( \alpha,\alpha^{\ast}\right) \Delta
\left( \alpha,\alpha^{\ast}\right) d^{2}\alpha,  \label{t17}
\end{equation}
where $W\left( q,p\right) =W\left( \alpha,\alpha^{\ast}\right) $ is the
Wigner function of density operator $\rho$, and the Wigner operator $\Delta
\left( q,p\right) $ is defined by
\begin{align}
\Delta \left( q,p\right) & =\frac{1}{2\pi}\int_{-\infty}^{\infty}dye^{iyp}%
\left \vert q+\frac{y}{2}\right \rangle \left \langle q-\frac{y}{2}\right
\vert  \notag \\
& =\frac{1}{\pi}\colon \exp \left \{ -(q-Q)^{2}-(p-P)^{2}\right \} \colon,
\label{t18}
\end{align}
or
\begin{align}
\Delta \left( q,p\right) & =\frac{1}{\pi}\colon \exp \left \{ -2(a^{\dagger
}-\alpha^{\ast})(a-\alpha)\right \} \colon  \label{t19} \\
& \equiv \Delta \left( \alpha,\alpha^{\ast}\right) ,  \notag
\end{align}
where the symbol $\colon \colon$ denotes the normal ordering.

Noticing that the transposed Wigner operator is given by
\begin{align}
\left[ \Delta \left( q,p\right) \right] ^{T} & =\Delta \left( q,-p\right) ,
\label{t20} \\
\left[ \Delta \left( \alpha,\alpha^{\ast}\right) \right] ^{T} & =\Delta
\left( \alpha^{\ast},\alpha \right) ,  \label{t21}
\end{align}
then the transposed density operator is obtained as%
\begin{align}
\rho^{T} & =2\pi \int_{-\infty}^{\infty}W\left( q,p\right) \Delta \left(
q,-p\right) dqdp  \notag \\
& =2\pi \int_{-\infty}^{\infty}W\left( q,-p\right) \Delta \left( q,p\right)
dqdp.  \label{t22}
\end{align}
Thus the Wigner function of transposed density operator $\rho^{T}$ can be
obtained by replacing $p$ with $-p$, i.e.,%
\begin{align}
W\left( q,p\right) & \rightarrow W\left( q,-p\right) ,  \notag \\
W\left( \alpha,\alpha^{\ast}\right) & \rightarrow W\left( \alpha^{\ast
},\alpha \right) .  \label{t23}
\end{align}
It is ready to see that the Wigner function of transposed density operator $%
\rho^{T}$ is still a real function which can be seen from Eq. (\ref{t29b}),
and $W\left( q,-p\right) $ is still a quasi-probability function because the
transpose can not change the trace of $\rho$, i.e.,
\begin{equation}
\text{tr}\rho^{T}=\int_{-\infty}^{\infty}W\left( q,-p\right) dqdp=1.
\label{t23b}
\end{equation}
It is interesting to notice that after the transpose, the Wigner function
and characteristic function can be obtained by mirror reflection along $q$
and $p$ axes, respectively \cite{2}. In addition, from Eq. (\ref{t22}) it is
clear that
\begin{align}
\left( \rho^{T}\right) ^{\dagger} & =2\pi \int_{-\infty}^{\infty}[W\left(
q,-p\right) ]^{\ast}\left[ \Delta \left( q,p\right) \right] ^{\dagger }dqdp
\notag \\
& =2\pi \int_{-\infty}^{\infty}W\left( q,-p\right) \Delta \left( q,p\right)
dqdp  \notag \\
& =\rho^{T}.  \label{t23c}
\end{align}
Eq. (\ref{t23c}) implies that $\rho^{T}$ is still an Hermite operator.
Noticing that the transpose can not change the eigenvalues, i.e., both $%
\rho^{T}$ and $\rho$ share the same non-negative real eigenvalues. Thus
\emph{for single-mode system, the transposed density operator is still a
density operator, which can represent a quantum state (eigenvalues of
density operator are non-negative real number)}.

\subsection{Average value of operator in the transposed density operator}

In this subsection, we consider the average value of operator under the
transposed density operator, and construct the relation between these
average values after and before the transpose operation.

\subsubsection{Average value in normally ordering form}

For any operator $\hat{O}$ whose normal ordering form is defined as $\colon
F\left( a,a^{\dag }\right) \colon $, then its average value under the
transposed density operator $\rho ^{T}$ can be calculated as%
\begin{equation}
\left\langle \colon F\left( a,a^{\dag }\right) \colon \right\rangle _{\rho
^{T}}=\left\langle \colon F\left( a^{\dag },a\right) \colon \right\rangle
_{\rho },  \label{t23d}
\end{equation}%
where the right-hand-side average value is under the density operator $\rho $%
. Eq. (\ref{t23d}) can be proved by extending $\rho $ in Fock space and
using the way similar to that deriving Eq. (\ref{t3c}). In particular, when
there is only real parameter within $\colon F\left( a^{\dag },a\right)
\colon $, then one can simply have $\left\langle \colon F\left( a,a^{\dag
}\right) \colon \right\rangle _{\rho ^{T}}=\left\langle \colon \lbrack
F\left( a,a^{\dag }\right) ]^{\dag }\colon \right\rangle _{\rho }$. These
above results can be extended directly to multimode transpose case.

\subsubsection{Covariance matrix of the transposed single-mode Gaussian state%
}

In the quantum system with continuous variable, covariance matrix is often
calculated, which consists of some average values. For example, covariance
matrix is used to derive the secure key rate for quantum key distribution
with continuous variable, to discuss the quantum steering of Gaussian state,
especially for investigating the entanglement of two-mode Gaussian state.
Here we take covariance matrix as an example, and construct the relation of
covariance matrix for Gaussian states before and after the transpose.

For single-mode Gaussian\ state, its characteristic function $\chi \left(
q,p\right) $ can be given by \cite{27}
\begin{equation}
\chi \left( \mathbf{\xi }\right) =\exp \left[ -\frac{1}{2}\mathbf{\xi }%
\Omega V\Omega ^{T}\mathbf{\xi }^{T}\mathbf{+}i\bar{X}^{T}\mathbf{\xi }^{T}%
\right] ,  \label{t24}
\end{equation}%
where $\bar{X}=$tr$\left( \hat{X}\rho \right) ,\mathbf{\xi =}\left(
\begin{array}{cc}
q & p%
\end{array}%
\right) $ and $V=\left(
\begin{array}{cc}
a_{11} & a_{12} \\
a_{21} & a_{22}%
\end{array}%
\right) $ is just the covariance matrix, as well as $\Omega =\left(
\begin{array}{cc}
0 & 1 \\
-1 & 0%
\end{array}%
\right) $. After the transpose, according to Eq. (\ref{t15}), then the
characteristic function of the transposed density operator is%
\begin{equation}
\chi _{\rho ^{T}}\left( \mathbf{\xi }\right) =\chi \left( -q,p\right) =\exp %
\left[ -\frac{1}{2}\mathbf{\xi }\Sigma \Omega V\Omega ^{T}\Sigma ^{T}\mathbf{%
\xi }^{T}\right] ,  \label{t25}
\end{equation}%
where $\Sigma =\left(
\begin{array}{cc}
-1 & 0 \\
0 & 1%
\end{array}%
\right) $ and we have assumed $\bar{X}=0$ for simplicity. Noticing that $%
\Omega \Omega ^{T}=\Omega ^{T}\Omega =1$, the covariance matrix of the
transposed density operator is given by%
\begin{align}
\Omega ^{T}\Sigma \Omega V\Omega ^{T}\Sigma ^{T}\Omega & =\allowbreak \left(
\begin{array}{cc}
1 & 0 \\
0 & -1%
\end{array}%
\right) V\left(
\begin{array}{cc}
1 & 0 \\
0 & -1%
\end{array}%
\right) ^{T}  \notag \\
& =\left(
\begin{array}{cc}
a_{11} & -a_{12} \\
-a_{21} & a_{22}%
\end{array}%
\right) .  \label{t26}
\end{align}%
Eq. (\ref{t26}) is just the relation between the two covariance matrixes
before and after the transpose.

In order to check the relation in Eq. (\ref{t26}), here we further consider
the special covariance matrix, which is defined by%
\begin{equation}
V=\left(
\begin{array}{cc}
\left \langle Q^{2}\right \rangle & \frac{1}{2}\left \langle QP+PQ\right
\rangle \\
\frac{1}{2}\left \langle QP+PQ\right \rangle & \left \langle P^{2}\right
\rangle%
\end{array}
\right) ,  \label{t27}
\end{equation}
where $Q=a+a^{\dag}$ and $P=(a-a^{\dag})/i$. Then we have
\begin{align}
Q^{2} & =a^{2}+a^{\dag2}+aa^{\dag}+a^{\dag}a  \notag \\
& =\colon a^{2}+a^{\dag2}+2a^{\dag}a+1\colon.  \label{t28}
\end{align}
According to Eq. (\ref{t23d}) we have
\begin{equation}
\left \langle Q^{2}\right \rangle _{\rho^{T}}=\left \langle \colon a^{\dag
2}+a^{2}+2a^{\dag}a+1\colon \right \rangle _{\rho}=\left \langle Q^{2}\right
\rangle _{\rho}.  \label{t28a}
\end{equation}
In a similar way, we have $\left \langle QP+PQ\right \rangle
_{\rho^{T}}=-\left \langle QP+PQ\right \rangle _{\rho},$ as expected.

\section{Two-mode operator's partial transpose}

In this section, we extend single-mode case to two-mode one. If we make the
transpose for two parties of two-mode quantum system, then the case is
similar to the above discussed, which will be a trivial extension. Here we
only consider partial transpose with respect to one part of two-mode, which
will be different from this case above in a certain extent and can also be
easily extended to multi-mode case.

\subsection{Partial transpose for any two-mode operator}

In Fock space, any two-mode operator $\hat{O}^{AB}$ can expanded as%
\begin{equation}
\hat{O}=\sum_{i,j,k,l}O_{i,jk,l}\left \vert i,j\right \rangle \left \langle
k,l\right \vert ,O_{i,jk,l}=\left \langle i,j\right \vert \hat{O}\left \vert
k,l\right \rangle ,  \label{t60}
\end{equation}
its partial transpose with respect to mode $B$, denoted as $\left( \hat {O}%
^{AB}\right) ^{T_{B}}$, can be defined as%
\begin{align}
\left( \hat{O}^{AB}\right) ^{T_{B}} & =\sum_{i,j,k,l}O_{i,jk,l}\left( \left
\vert i,j\right \rangle \left \langle k,l\right \vert \right) ^{T_{B}}
\notag \\
& =\sum_{i,j,k,l}O_{i,jk,l}\left \vert i,l\right \rangle \left \langle
k,j\right \vert .  \label{t61}
\end{align}
In a similar way to derive Eq. (\ref{t3a}), for two-mode normally ordered
operator $\colon F\left( a,a^{\dag};b,b^{\dag}\right) \colon$, its partial
transpose for mode $B$ is given by%
\begin{equation}
\left[ \colon F\left( a,a^{\dag};b,b^{\dag}\right) \colon \right]
^{T_{B}}=\colon F\left( a,a^{\dag};b^{\dag},b\right) \colon.  \label{t62}
\end{equation}
This shows that the partial transpose for operator such as $\colon F\left(
a,a^{\dag};b,b^{\dag}\right) \colon$\ can be achieved by replacing $%
b(b^{\dag})$ with $b^{\dag}(b)$ respectively, while the other mode $A$ is
kept unchanged. Here we should emphasize that the replacement must be taken
under the normal ordering form. This point will become clear in the
following.

\subsection{Partially transposed density operator in phase space}

Next, considering any two-mode Bose density operator $\rho_{AB}$, it can
always be expanded as%
\begin{align}
\rho_{AB} & =\int_{-\infty}^{\infty}\frac{dq_{1}dp_{1}dq_{2}dp_{2}}{\left(
2\pi \right) ^{2}}\chi \left( q_{1},p_{1};q_{2},p_{2}\right)  \notag \\
& \times D_{a}\left( -q_{1},-p_{1}\right) D_{b}\left( -q_{2},-p_{2}\right) ,
\label{t31}
\end{align}
or%
\begin{align}
\rho_{AB} & =\int_{-\infty}^{\infty}dq_{1}dp_{1}dq_{2}dp_{2}W\left(
q_{1},p_{1};q_{2},p_{2}\right)  \notag \\
& \times \left( 2\pi \right) ^{2}\Delta_{a}\left( q_{1},p_{1}\right)
\Delta_{b}\left( q_{2},p_{2}\right) .  \label{t32}
\end{align}
Here we take the partial transpose with respect to mode $B$. Using Eqs. (\ref%
{t10}) and (\ref{t20}) it is ready to have
\begin{align}
\chi \left( q_{1},p_{1};q_{2},p_{2}\right) & \rightarrow \chi \left(
q_{1},p_{1};-q_{2},p_{2}\right) ,  \notag \\
W\left( q_{1},p_{1};q_{2},p_{2}\right) & \rightarrow W\left(
q_{1},p_{1};q_{2},-p_{2}\right) ,  \label{t33}
\end{align}
then we have%
\begin{align}
\left( \rho_{AB}\right) ^{T_{B}} &
=\int_{-\infty}^{\infty}dq_{1}dp_{1}dq_{2}dp_{2}W\left(
q_{1},p_{1};q_{2},-p_{2}\right)  \notag \\
& \times \left( 2\pi \right) ^{2}\Delta_{a}\left( q_{1},p_{1}\right)
\Delta_{b}\left( q_{2},p_{2}\right) ,  \label{t34}
\end{align}
and%
\begin{align}
\left( \rho_{AB}\right) ^{T_{B}} & =\int_{-\infty}^{\infty}\frac {%
dq_{1}dp_{1}dq_{2}dp_{2}}{\left( 2\pi \right) ^{2}}\chi \left(
q_{1},p_{1};-q_{2},p_{2}\right)  \notag \\
& \times D_{a}\left( -q_{1},-p_{1}\right) D_{b}\left( -q_{2},-p_{2}\right) .
\label{t35}
\end{align}
In a similar way, one can see that
\begin{equation}
\left[ \left( \rho_{AB}\right) ^{T_{B}}\right] ^{\dagger}=\left(
\rho_{AB}\right) ^{T_{B}},  \label{t36}
\end{equation}
and
\begin{align}
1 & =\text{tr}\left( \rho_{AB}\right) ^{T_{B}}  \notag \\
& =\int_{-\infty}^{\infty}dq_{1}dp_{1}dq_{2}dp_{2}W\left(
q_{1},p_{1};q_{2},-p_{2}\right) .  \label{t37}
\end{align}
Here we should emphasize that although the partially transposed density
operator $\left( \rho_{AB}\right) ^{T_{B}}$ presents the properties of both
quasi-probability and Hermite, $\left( \rho_{AB}\right) ^{T_{B}}$ and $%
\rho_{AB}$ generally will not share the same non-negative real eigenvalues,
which implies that for two-mode system, the partially transposed density
operator may not be qualified as a density operator.

\subsection{Partial transpose of two-mode squeezing operator}

Here we examine the partial transpose of two-mode squeezing operator as an
example. Theoretically, the two-mode squeezing operator is expressed as \cite%
{28}
\begin{equation}
S\left( \xi \right) =\exp \left\{ \left( \xi a^{\dagger }b^{\dagger }-\xi
^{\ast }ab\right) \right\} ,\xi =re^{i\varphi }  \label{t42}
\end{equation}%
whose normal ordering form is given by \cite{25}%
\begin{align}
S\left( \xi \right) & =\text{sech}r\colon \exp \left\{ a^{\dagger
}b^{\dagger }e^{i\varphi }\tanh r\right\}  \notag \\
& \times \exp \left\{ \left( a^{\dagger }a+b^{\dagger }b\right) \left( \text{%
sech}r-1\right) \right\}  \notag \\
& \times \exp \left\{ -abe^{-i\varphi }\tanh r\right\} \colon .  \label{t43}
\end{align}%
In a similar way to derive Eq. (\ref{t28b}), one can get
\begin{align}
\left[ S\left( \xi \right) \right] ^{T_{B}}& =\text{sech}r\colon \exp
\left\{ a^{\dagger }be^{i\varphi }\tanh r\right\}  \notag \\
& \times \exp \left\{ \left( a^{\dagger }a+b^{\dagger }b\right) \left( \text{%
sech}r-1\right) \right\}  \notag \\
& \times \exp \left\{ -ab^{\dagger }e^{-i\varphi }\tanh r\right\} \colon .
\label{t44}
\end{align}%
On the other hand, for a beam splitter operator, we have \cite{29,30}
\begin{align}
B(\theta )& =\exp \left\{ \theta \left( a^{\dagger }be^{i\varphi
}-ab^{\dagger }e^{-i\varphi }\right) \right\}  \notag \\
& =\colon \exp \left\{ \left( a^{\dagger }a+b^{\dagger }b\right) \left( \cos
\theta -1\right) \right\}  \notag \\
& \times \exp \left\{ \left( a^{\dagger }be^{i\varphi }-ab^{\dagger
}e^{-i\varphi }\right) \sin \theta \right\} .  \label{t45}
\end{align}%
Comparing Eqs. (\ref{t44}) with (\ref{t45}), it is interesting to notice
that if we take $\cos \theta =$sech$r,\sin \theta =\tanh r,\theta =\arg \sin %
\left[ \tanh r\right] ,$ then Eq. (\ref{t44}) can be reformed as
\begin{align}
\left[ S\left( \xi \right) \right] ^{T_{B}}& =\cos \theta \exp \left\{
\theta \left( a^{\dagger }be^{i\varphi }-ab^{\dagger }e^{-i\varphi }\right)
\right\}  \notag \\
& =\cos \theta \times B(\theta ).  \label{t47}
\end{align}%
It is shown that the partial transpose of two-mode squeezing operator can be
equivalent to a beam splitter operator except for a factor $\cos \theta $.
Here, we should emphasize that the partial transpose of two-mode squeezing
operator cannot be obtained by simply replacing $b(b^{\dagger })$ with $%
b^{\dagger }(b)$ from Eq. (\ref{t42}). This operation is valid only within
the normal ordering form.

\subsection{Covariance matrix of partially transposed two-mode Gaussian
states}

For two-mode Gaussian states, the characteristic function of density
operator $\rho_{AB}$ can be shown as \cite{27}

\begin{equation}
\chi \left( \mathbf{\xi}\right) =\exp \left[ -\frac{1}{2}\mathbf{\xi}\Omega
V\Omega^{T}\mathbf{\xi}^{T}\right] ,  \label{t48}
\end{equation}
where we have assumed $\bar{X}=0$ again for simplicity, and $\bar{X}=$tr$%
\left( \hat{X}\rho \right) ,\mathbf{\xi=}\left(
\begin{array}{cccc}
q_{1} & p_{1} & q_{2} & p_{2}%
\end{array}
\right) $ and the covariance matrix $V$ and $\Omega$, respectively, are
given by%
\begin{align}
V & =\left(
\begin{array}{cccc}
a_{11} & a_{12} & a_{13} & a_{14} \\
a_{21} & a_{22} & a_{23} & a_{24} \\
a_{31} & a_{32} & a_{33} & a_{34} \\
a_{41} & a_{42} & a_{43} & a_{44}%
\end{array}
\right) =V^{T},  \label{t49} \\
\Omega & =\left(
\begin{array}{cccc}
0 & 1 & 0 & 0 \\
-1 & 0 & 0 & 0 \\
0 & 0 & 0 & 1 \\
0 & 0 & -1 & 0%
\end{array}
\right) ,\Omega \Omega^{T}=1.  \label{t50}
\end{align}
According to Eq. (\ref{t33}), the characteristic function of partially
transposed density operator $\left( \rho_{AB}\right) ^{T_{B}}$ is given by $%
\chi_{\left( \rho_{AB}\right) ^{T_{B}}}\left( \mathbf{\xi}\right) =\exp %
\left[ -\frac{1}{2}\mathbf{\xi}\Sigma \Omega V\Omega^{T}\Sigma ^{T}\mathbf{%
\xi}^{T}\right] ,$ with $\Sigma=\left(
\begin{array}{cccc}
1 & 0 & 0 & 0 \\
0 & 1 & 0 & 0 \\
0 & 0 & -1 & 0 \\
0 & 0 & 0 & 1%
\end{array}
\right) .$ Thus the covariance matrix of partially transposed density
operator is given by%
\begin{align}
V_{\left( \rho_{AB}\right) ^{T_{B}}} & =\Omega^{T}\Sigma \Omega V\Omega
^{T}\Sigma^{T}\Omega=\bar{\Sigma}V\bar{\Sigma}^{T}  \notag \\
& =\allowbreak \left(
\begin{array}{cccc}
a_{11} & a_{12} & a_{13} & -a_{14} \\
a_{21} & a_{22} & a_{23} & -a_{24} \\
a_{31} & a_{32} & a_{33} & -a_{34} \\
-a_{41} & -a_{42} & -a_{43} & a_{44}%
\end{array}
\right) ,  \label{t51}
\end{align}
with $\bar{\Sigma}=\left(
\begin{array}{cccc}
1 & 0 & 0 & 0 \\
0 & 1 & 0 & 0 \\
0 & 0 & 1 & 0 \\
0 & 0 & 0 & -1%
\end{array}
\right) .$ Eq. (\ref{t51}) is just the relation between the two covariance
matrixes before and after the partial transpose. This case can be easily
extended to multi-mode case.

In order to check the relation in Eq. (\ref{t51}), we consider the average
value of $a_{14}=\left \langle Q_{1}P_{2}\right \rangle $ where $%
Q_{1}=(a_{1}+a_{1}^{\dag})$ and $P_{2}=(a_{2}-a_{2}^{\dag})/i$. It is ready
to see $Q_{1}P_{2}=\colon(a_{1}+a_{1}^{\dag})(a_{2}-a_{2}^{\dag})/i\colon.$
Using Eq. (\ref{t23d}), we have%
\begin{align}
\left \langle Q_{1}P_{2}\right \rangle _{\left( \rho_{AB}\right) ^{T_{B}}} &
=\left \langle \colon(a_{1}+a_{1}^{\dag})(a_{2}^{\dag}-a_{2})/i\colon \right
\rangle _{\rho_{AB}}  \notag \\
& =-\left \langle Q_{1}P_{2}\right \rangle _{\rho_{AB}},  \label{t52}
\end{align}
as expected.

\section{Entanglement of two-mode squeezed states in laser channels}

In this section, we further consider another important application of Eq. (%
\ref{t62}). Actually, the partial transpose is usually used to examine
whether there is entanglement for multi-mode quantum system after one or
more modes\ are transposed while others are kept unchanged. Here we take
two-mode case as an example, and derive analytically the entanglement degree
in laser channels. For this purpose, we first briefly review the logarithmic
negativity, then derive analytical expressions of both normal ordering form
of density operator and its logarithmic negativity in the laser channels.

\subsection{Negativity and logarithmic negativity}

If a two-mode quantum state is separable, then its density operator can be
expressed as
\begin{equation}
\rho _{AB}=\sum_{i}p_{i}\rho _{A_{i}}\otimes \rho _{B_{i}},  \label{t38}
\end{equation}%
thus the corresponding partial transpose with respect to $B$ mode is given by%
\begin{equation}
\left( \rho _{AB}\right) ^{T_{B}}=\sum_{i}p_{i}\rho _{A_{i}}\otimes \left(
\rho _{B_{i}}\right) ^{T}.  \label{t39}
\end{equation}%
As mentioned above, $\left( \rho _{B_{i}}\right) ^{T}$ is still a quantum
state. Thus for a separable state, the partially transposed operator in Eq. (%
\ref{t39}) can be qualified as a quantum state. Non-negative eigenvalues are
shared by the two states after and before the partial transpose. However,
for an entangled state $\rho _{AB},$ it can not be expressed as the form in
Eq. (\ref{t38}), thus its partial transpose may not be a quantum state
although the partially transposed operator satisfies both Eqs. (\ref{t36})
and (\ref{t37}). This implies that there may have negative eigenvalues for $%
\left( \rho _{AB}\right) ^{T_{B}}$. Based on this point, one can quantify
the degree of entanglement by using negative eigenvalues \cite{11}, i.e., by
defining
\begin{align}
N\left( \rho _{AB}\right) & =\frac{1}{2}\text{Tr}\left[ \sqrt{[\left( \rho
_{AB}\right) ^{T_{B}}]^{2}}-\left( \rho _{AB}\right) ^{T_{B}}\right]  \notag
\\
& =\frac{\left\Vert \left( \rho _{AB}\right) ^{T_{B}}\right\Vert -1}{2},
\label{t40}
\end{align}%
and
\begin{equation}
E\left( \rho _{AB}\right) =\log _{2}\left[ 1+2N\left( \rho _{AB}\right) %
\right] =\log _{2}\left\Vert \left( \rho _{AB}\right) ^{T_{B}}\right\Vert ,
\label{t41}
\end{equation}%
where $\left\Vert \hat{O}\right\Vert $ is the trace norm, $\left\Vert \hat{O}%
\right\Vert =\sqrt{\hat{O}^{\dagger }\hat{O}}.$ Eqs. (\ref{t40}) and (\ref%
{t41}) are the negativity and the logarithmic negativity, respectively. Both
of them are effective for Gaussian states and non-Gaussian states and are
much more amenable to evaluation than distillable entanglement, relative
entropy entanglement and entanglement cost.

\subsection{Normal ordering form of density operator in the laser channel}

Next, we consider the evolution of density operator in laser channel, which
includes photon loss, thermal channel as its special cases. For single-mode
system, the master equation describing the dynamical process is given by
\cite{14}%
\begin{align}
\frac{d}{dt}\rho \left( t\right) & =g\left[ 2a^{\dag}\rho
a-aa^{\dag}\rho-\rho aa^{\dag}\right]  \notag \\
& +\kappa \left[ 2a\rho a^{\dag}-a^{\dag}a\rho-\rho a^{\dag}a\right] ,
\label{r1}
\end{align}
where $g$ and $\kappa$ are the gain and the loss factors, respectively.

By using the entangled state representation, from Eq. (\ref{r1}) one can
obtain the evolution of Wigner function as \cite{16}%
\begin{equation}
W\left( \alpha ,t\right) =\frac{2}{A}\int \frac{d^{2}\beta }{\pi }e^{-\frac{2%
}{A}\left\vert \alpha -\beta e^{-\left( \kappa -g\right) t}\right\vert
^{2}}W\left( \beta ,0\right) ,  \label{r2}
\end{equation}%
where $W\left( \beta ,0\right) $ is the initial Wigner function and
\begin{equation*}
A=\frac{\kappa +g}{\kappa -g}\left( 1-e^{-2\left( \kappa -g\right) t}\right)
.
\end{equation*}%
For continuous variable system, two-mode entangled Gaussian states are main
resource of quantum information and processing and easily generated
experimentally. Thus many researchers have paid much attention to them \cite%
{31,32,33,34,35}. Here, we consider two-mode of the TMSV passing the
channels, independently. The initial Wigner function of the TMSV is given by%
\begin{align}
W\left( \beta _{1},\beta _{2},0\right) & =\frac{1}{\pi ^{2}}\exp
\{-2(\allowbreak \left\vert \beta _{1}\right\vert ^{2}+\left\vert \beta
_{2}\right\vert ^{2})\cosh 2r\}  \notag \\
& \times \exp \left\{ 2\left( \beta _{1}\beta _{2}+\beta _{1}^{\ast
}\allowbreak \beta _{2}^{\ast }\right) \sinh 2r\right\} .  \label{r4}
\end{align}%
Substituting Eq. (\ref{r4}) into Eq. (\ref{r2}) one can get the Wigner
function after the laser channels, i.e.,%
\begin{align}
W\left( \alpha _{1},\alpha _{2},t\right) & =\frac{2^{2}}{A_{1}A_{2}}\int
\frac{d^{2}\beta _{1}d^{2}\beta _{2}}{\pi ^{2}}W\left( \beta _{1},\beta
_{2},0\right)  \notag \\
& \times \exp \left[ -\frac{2}{A_{1}}\left\vert \alpha _{1}-\beta
_{1}e^{-\left( \kappa _{1}-g_{1}\right) t}\right\vert ^{2}\right]  \notag \\
& \times \exp \left[ -\frac{2}{A_{2}}\left\vert \alpha _{2}-\beta
_{2}e^{-\left( \kappa _{2}-g_{2}\right) t}\right\vert ^{2}\right] ,
\label{r5}
\end{align}%
where $g_{l}$ and $\kappa _{l}(l=1,2)$ are the gain and the loss factors,
respectively, and
\begin{equation}
A_{j}=\frac{\kappa _{j}+g_{j}}{\kappa _{j}-g_{j}}\left( 1-e^{-2(\kappa
_{j}-g_{j})t}\right) ,j=1,2.  \label{r6}
\end{equation}%
From Eq. (\ref{r5}), it is ready to obtain%
\begin{align}
W\left( \alpha _{1},\alpha _{2},t\right) & =\frac{1}{\pi ^{2}\Gamma }\exp
\left\{ -2\lambda _{1}\left\vert \alpha _{1}\right\vert ^{2}-2\lambda
_{2}\left\vert \alpha _{2}\right\vert ^{2}\right\}  \notag \\
& \exp \left\{ 2\lambda _{12}\left( \alpha _{2}^{\ast }\alpha _{1}^{\ast
}+\alpha _{1}\alpha _{2}\right) \right\} ,  \label{r7}
\end{align}%
where we have set%
\begin{align}
\lambda _{1}& =\frac{1}{\Gamma }\left( A_{2}+e^{-2\left( \kappa
_{2}-g_{2}\right) t}\cosh 2r\right) ,  \notag \\
\lambda _{2}& =\frac{1}{\Gamma }\left( A_{1}+e^{-2\left( \kappa
_{1}-g_{1}\right) t}\cosh 2r\right) ,  \notag \\
\lambda _{12}& =\frac{1}{\Gamma }e^{-\left( \kappa _{1}-g_{1}\right)
t}e^{-\left( \kappa _{2}-g_{2}\right) t}\sinh 2r,  \label{r10}
\end{align}%
and%
\begin{align}
\Gamma & =\left( A_{1}e^{-2\left( \kappa _{2}-g_{2}\right)
t}+A_{2}e^{-2\left( \kappa _{1}-g_{1}\right) t}\right) \cosh 2r  \notag \\
& +A_{1}A_{2}+e^{-2\left( \kappa _{1}-g_{1}\right) t}e^{-2\left( \kappa
_{2}-g_{2}\right) t}.  \label{r11}
\end{align}%
Then using Eq. (\ref{t17}) or Eq. (\ref{t32}) and the integration within an
ordered product of operators (IWOP) technique \cite{36,37}, one can get the
final density operator by performing the integration, i.e.,
\begin{align}
\rho _{ab}& =\left( 4\pi \right) ^{2}\int W\left( \alpha _{1},\alpha
_{2},t\right) \Delta _{a}\left( \alpha _{1}\right) \Delta _{b}\left( \alpha
_{2}\right) d^{2}\alpha _{1}d^{2}\alpha _{2}  \notag \\
& =4\Omega _{0}\colon e^{2\Omega _{1}a^{\dagger }a+2\Omega _{2}b^{\dagger
}b+2\Omega _{3}\left( ab+a^{\dagger }b^{\dagger }\right) }\colon ,
\label{r12}
\end{align}%
where
\begin{align}
\Omega _{0}& =\frac{1/\Gamma }{\left( \lambda _{2}+1\right) \left( \lambda
_{1}+1\right) -\lambda _{12}^{2}}>0,  \notag \\
\Omega _{1}& =\frac{\lambda _{12}^{2}-\lambda _{1}-\lambda _{1}\lambda _{2}}{%
\left( \lambda _{2}+1\right) \left( \lambda _{1}+1\right) -\lambda _{12}^{2}}%
,  \notag \\
\Omega _{2}& =\frac{\lambda _{12}^{2}-\lambda _{2}-\lambda _{1}\lambda _{2}}{%
\left( \lambda _{2}+1\right) \left( \lambda _{1}+1\right) -\lambda _{12}^{2}}%
,  \notag \\
\Omega _{3}& =\frac{\lambda _{12}}{\left( \lambda _{2}+1\right) \left(
\lambda _{1}+1\right) -\lambda _{12}^{2}}.  \label{r13}
\end{align}%
Eq. (\ref{r12}) is just the normal ordering form of the density operator
after the laser channels.

\subsection{Trace norm of decohered TMSV in the laser channel}

According to Eq. (\ref{t3a}), it is ready to get the partial transpose for
party $b$ of $\rho _{ab},$ i.e., replacing $b^{\dagger }\ $with $b$, then%
\begin{align}
\left[ \rho _{ab}\right] ^{T_{b}}& =4\Omega _{0}\colon e^{2\Omega
_{1}a^{\dagger }a+2\Omega _{2}b^{\dagger }b+2\Omega _{3}\left( ab^{\dagger
}+a^{\dagger }b\right) }\colon  \notag \\
& =4\Omega _{0}\colon e^{2\left(
\begin{array}{cc}
a^{\dagger } & b^{\dagger }%
\end{array}%
\right) \left(
\begin{array}{cc}
\Omega _{1} & \Omega _{3} \\
\Omega _{3} & \Omega _{2}%
\end{array}%
\right) \left(
\begin{array}{c}
a \\
b%
\end{array}%
\right) }\colon .  \label{r14}
\end{align}%
Then using the completeness of two-mode coherent states or the following
formula \cite{38,38b}
\begin{align}
& \left\{ \colon \exp [\mathbf{\tilde{a}}^{\dag }\left( u-I\right) \mathbf{a}%
]\colon \right\} \left\{ \colon \exp [\mathbf{\tilde{a}}^{\dag }\left(
u^{\prime }-I\right) \mathbf{a}]\colon \right\}  \notag \\
& =\colon \exp [\mathbf{\tilde{a}}^{\dag }\left( uu^{\prime }-I\right)
\mathbf{a}]\colon ,  \label{r14b}
\end{align}%
where $\mathbf{a=}\left(
\begin{array}{cccc}
a & b & \cdots & d%
\end{array}%
\right) ^{T}$ and $\mathbf{\tilde{a}}^{\dag }=\left(
\begin{array}{cccc}
a^{\dag } & b^{\dag } & \cdots & d^{\dag }%
\end{array}%
\right) $, the squared of \{$[\rho _{ab}]^{T_{b}}\}^{2}$ can be calculated as%
\begin{equation}
\{[\rho _{ab}]^{T_{b}}\}^{2}=\left( 4\Omega _{0}\right) ^{2}\colon \exp
\left\{ \left(
\begin{array}{cc}
a^{\dagger } & b^{\dagger }%
\end{array}%
\right) \left[ M-I\right] \left(
\begin{array}{c}
a \\
b%
\end{array}%
\right) \right\} \colon ,  \label{r15}
\end{equation}%
where we have set $\allowbreak $
\begin{align*}
M& =\left(
\begin{array}{cc}
w & v \\
v & u%
\end{array}%
\right) , \\
w& =\left( 1+2\Omega _{1}\right) ^{2}+\left( 2\Omega _{3}\right) ^{2}, \\
u& =\left( 1+2\Omega _{2}\right) ^{2}+\left( 2\Omega _{3}\right) ^{2}, \\
v& =4\Omega _{3}\left( 1+\Omega _{1}+\Omega _{2}\right) .
\end{align*}%
It is clear that $w>0,$ $u>0$ and $wu-v^{2}\geqslant 0$. Then using the
formula converting normal ordering form of operator to its close exponential
form \cite{38,38b}%
\begin{equation}
\hat{O}\equiv \colon \exp \left\{ \mathbf{\tilde{a}}^{\dag }\left(
N-I\right) \mathbf{a}\right\} \colon =\exp \left\{ \mathbf{\tilde{a}}^{\dag
}\left( \ln N\right) \mathbf{a}\right\} ,  \label{r17}
\end{equation}%
which leads to
\begin{equation}
\sqrt{\hat{O}}=\colon \exp \left\{ \left(
\begin{array}{cc}
a^{\dagger } & b^{\dagger }%
\end{array}%
\right) \left( \sqrt{N}-I\right) \left(
\begin{array}{c}
a \\
b%
\end{array}%
\right) \right\} \colon ,  \label{r17b}
\end{equation}%
then we have
\begin{align}
& \sqrt{\{[\rho _{ab}]^{T_{b}}\}^{2}}  \notag \\
& =4\Omega _{0}\colon \exp \left\{ \left(
\begin{array}{cc}
a^{\dagger } & b^{\dagger }%
\end{array}%
\right) \left( \sqrt{M}-I\right) \left(
\begin{array}{c}
a \\
b%
\end{array}%
\right) \right\} \colon ,  \label{r18}
\end{align}%
where we may define $\sqrt{M}=\left(
\begin{array}{cc}
p & m \\
m & q%
\end{array}%
\right) $ due to the symmetry of $M$.

Further using the completeness of coherent state representation again, one
can get
\begin{equation}
\left\Vert \left[ \rho _{ab}\right] ^{T_{b}}\right\Vert =\frac{4\Omega _{0}}{%
\det (I-\sqrt{M})}=\frac{4\Omega _{0}}{\Delta }.  \label{r20}
\end{equation}%
where $\Delta =\left( 1-p\right) \left( 1-q\right) -m^{2}=1+\sqrt{wu-v^{2}}%
-\Lambda _{+}-\Lambda _{-},$ and $\Lambda _{\pm }=\frac{1}{\sqrt{2}}\sqrt{%
w+u\pm \sqrt{4v^{2}+(w-u)^{2}}}.$ In particular, the condition $\det
M=wu-v^{2}\geqslant 0$ must be satisfied, which also leads to that $\Lambda
_{\pm }$ are positive real numbers. Then the logarithmic negativity of
decohered TMSV in the laser channel is given by%
\begin{equation}
E_{N}=\log _{2}\frac{4\Omega _{0}}{\Delta }.  \label{r21}
\end{equation}%
It is obvious that there is non-zero degree of entanglement only when $%
4\Omega _{0}>\Delta .$ This condition will determine the critical time or
the minimum disentanglement time.

\subsection{Some special cases}

Next, we pay attention to some special cases, including the ideal TMSV, the
TMSV decohered by bi-symmetrical photon-loss as well as gain and loss.

\emph{Entanglement of the TMSV: }For this case, we need to take $t=0$ which
leads to $2\Omega_{1}=2\Omega_{2}=-1,2\Omega_{3}=\tanh r,$ $4\Omega
_{0}=1/\cosh^{2}r$ and $w=u=\tanh^{2}r,v=0$ as well as $\Lambda_{\pm}=\tanh
r $, thus the logarithmic negativity of the TMSV is given by $E_{N}=\log
_{2}e^{2r}$, as expected.

\emph{Entanglement of the TMSV with bi-symmetrical photon-loss:} For this
case, we can take $g_{i}=0$ and $\kappa _{1}=\kappa _{2}=\kappa .$ Then we
have
\begin{align}
w& =u=\frac{T^{2}\left( 1+R^{2}\tanh ^{2}r\right) }{\left( 1-R^{2}\tanh
^{2}r\right) ^{2}}\tanh ^{2}r,  \label{r24a} \\
v& =\frac{2RT^{2}\tanh ^{3}r}{\left( 1-R^{2}\tanh ^{2}r\right) ^{2}},
\label{r24b}
\end{align}%
thus
\begin{equation}
E_{N}=\log _{2}\frac{1}{1-T\left( 1-e^{-2r}\right) },  \label{r24}
\end{equation}%
where $R=1-T$ and $T=e^{-2\kappa t}$. In particular, when $T=1$ Eq. (\ref%
{r24}) just reduce to the negativity $E_{N}=\log _{2}e^{2r}$ of the TMSV.
Eq. (\ref{r24}) is agreement with the result in Ref. \cite{34,39}, as
expected. It is easy to see that $E_{N}$ increases and decreases as the
increasing $r$ and $1-T$, respectively. In the limit of $T\rightarrow 0$, we
have $E_{N}\rightarrow 0$. This case is true for single-side photon-loss
case, in which $w=T\tanh ^{2}r,$ $v=\sqrt{T}R\tanh ^{3}r,$ and $u=\left(
R^{2}\tanh ^{2}r+T\right) \tanh ^{2}r$, and $E_{N}$ can be obtained by
substituting these into Eqs. (\ref{r20})-(\ref{r21}), not shown here for
simplicity. This indicates that the entanglement always exists only when the
loss is present.

\emph{Bi-symmetrical photon-loss and gain}: For this case, we can take $%
g_{i}=g,$ $\kappa _{i}=\kappa .$ From Eq. (\ref{r21}) we have
\begin{equation}
t_{c}=\frac{1}{2\left( \kappa -g\right) }\ln \frac{g+\kappa \tanh r}{g\left(
1+\tanh r\right) },  \label{r25}
\end{equation}%
for any squeezing parameter $r>0$, when $t\geqslant t_{c}$ then $E_{N}=0.$
It is shown that there exists a threshold value of time $t_{c}$ when the
system passes a channel with gain and loss. This threshold time is different
from the case only with loss.

\section{Conclusion}

In this paper, we examined both the transpose of single-mode operator and
the partial transpose of two-mode operator. Based on the definition of the
operator transpose in the Fock space and the normal ordering form of
operator, we derived a lemma, i.e., the transposed operator for both
single-mode and two-mode cases can be obtained by replacing transposed mode,
with respect to $b(b^{\dag })$ with $b^{\dag }(b)$, respectively, within
normal ordering form. Those classical numbers involved are kept unchanged.
Using this property, we further considered some transposed operators, such
as displaced operator, Wigner operator, single/two-mode squeezing operator.
Then we constructed the relation between distribution functions in phase
space before and after (partial) transpose, including Wigner function and
characteristic function. The relation of average values is also bridged,
including covariance matrix. These above discussions can be extended to
multimode case. As an effective and direct application of the lemma, we
analytically calculated the entanglement degree of the TMSV when passing
through the laser channel, by using the logarithmic negativity. In addition,
it will be effective for calculating entanglement of non-Gaussian states
using the normal ordering form of partially transposed density operator. Our
method could be effective for entangled Fermi ststems, which will be further
condidered in the future.

\section*{Acknowledgments}

This work is supported by the National Natural Science Foundation of China
(Grant Nos. 11964013,11664017), the Training Program for Academic and
Technical Leaders of Major Disciplines in Jiangxi Province (20204BCJL22053).

\section*{Disclosures}

The authors declare no conflicts of interest.

\end{document}